\documentclass[twocolumn,preprintnumbers,amsmath,amssymb,prb]{revtex4}

\usepackage{graphicx}
\usepackage{dcolumn}
\usepackage{bm}
\usepackage{amssymb}
\usepackage{epstopdf}
\usepackage{color}

\begin{document}

\title{All-optical control of excitons in semiconductor quantum wells}

\author{V. M. Kovalev}
\author{M. V. Boev}
\author{O. V. Kibis}\email{Oleg.Kibis@nstu.ru}
\affiliation {Department of Applied and Theoretical Physics, Novosibirsk~State~Technical~University,
Karl~Marx~Avenue~20,~Novosibirsk~630073,~Russia}

\begin{abstract}
Applying the Floquet theory, we developed the method to control excitonic properties of semiconductor quantum wells by a high-frequency electromagnetic field. It is demonstrated, particularly, that the field induces the blue shift of exciton emission from the quantum wells and narrows width of the corresponding spectral line. As a consequence, the field strongly modifies optical properties of the quantum wells and, therefore, can be used to tune characteristics of the optoelectronic devices based on them.
\end{abstract}

\maketitle

\section{Introduction}
The all-optical control of electronic properties of solids by a high-frequency off-resonant electromagnetic field, based ideologically on the Floquet theory of periodically driven quantum systems (Floquet engineering), remains an exciting field of modern physics during long time since it results in many light-induced effect in various nanostructures~\cite{Oka_2019,Basov_2017,Eckardt_2015,Goldman_2014,Bukov_2015,Casas_2001,Kibis_2020_1,Koshelev_2015,Rechtsman_2013,Wang_2013,Torres_2014,Calvo_2015,Mikami_2016,Oka_2009,Iurov_2017,Iurov_2013,Syzranov_2013,Usaj_2014,Perez_2014,Glazov_2014,Sentef_2015,Sie_2015,Kibis_2017,Iurov_2019,Iurov_2020,Cavalleri_2020}. Among the actively studied nanostructures, the semiconductor quantum wells (QWs)~\cite{Nag_2002book} take deserved place as building blocks of modern nanoelectronics. In previous studies, the Floquet theory was applied to irradiated QWs to describe the electron transport~\cite{Morina_2015}, the spin effects~\cite{Pervishko_2015}, the magnetic properties of electrons~\cite{Dini_2016}, polaritons~\cite{Kyriienko_2017} and the topological states~\cite{Lindner_2011}. However, the Floquet theory of excitons in QWs still wait for detailed analysis. This problem is of current importance since the excitons --- together with the electron scattering and capture processes~\cite{Zakhleniuk_1999,Stavrou_2001} --- form the main mechanism for light emission from semiconductor structures, which dominates for low temperatures~\cite{Ivchenko_2007book}. Correspondingly, the Floquet engineering of excitons will open the novel way to control radiative characteristics of various light-emitting optoelectronic devices based in the QWs~\cite{Nag_2002book,Zory_1993book}. The present article is aimed to fill partially this gap in the theory.

\section{Model}
\begin{figure}[h!]
\centering\includegraphics[width=1.0\columnwidth]{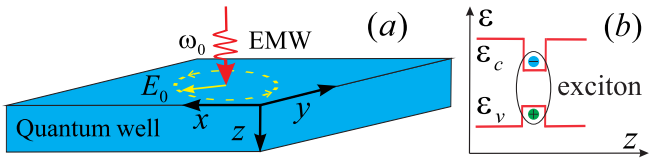}
\caption{Sketch of the system under consideration: (a) The semiconductor quantum well (QW) irradiated by the circularly polarized electromagnetic wave (EMW) with the frequency $\omega_0$ and the electric field amplitude $E_0$; (b) The energy band structure of the QW containing the electron-hole pair (exciton), where $\varepsilon_{c(v)}$ is the band edge of the conduction (valence) band of the semiconductor.} \label{fig1}
\end{figure}
Let us consider an exciton in a semiconductor QW irradiated by a circularly polarized electromagnetic wave (EMW) with the frequency $\omega_0$ and the electric field amplitude $E_0$, which propagates perpendicularly to the QW plane $x,y$ (see Fig.~1). In the following, the field frequency $\omega_0$ will be assumed to be below the optical absorption threshold of the QW. Therefore, the EMW can be treated as an off-resonant field which cannot be absorbed by charge carriers and  only ``dresses'' them (the dressing field), driving periodically electrons and holes. Correspondingly, the exciton Hamiltonian for the irradiated QW reads
\begin{equation}\label{HEMW}
\hat{\cal H}_{X}=\sum_{j=e,h}\frac{[\hat{\mathbf{p}}_j-e\xi_j\mathbf{A}(t)/c]^2}{2m_j}+U_C(\mathbf{r}_e,\mathbf{r}_h),
\end{equation}
where $\hat{p}_{e(h)}$ is the in-plane electron (hole) momentum operator, $m_{e(h)}$ is the electron (hole) effective mass,
\begin{equation}\label{A}
\mathbf{A}(t)=(A_x,A_y)=[cE_0/\omega_0](\sin\omega_0 t,\cos\omega_0 t)
\end{equation}
is the vector potential of the dressing field, $\xi_e=1$ and $\xi_h=-1$, $\mathbf{r}_{e(h)}$ is the in-plane radius-vector of electron (hole),
\begin{equation}\label{UC}
U_C(\mathbf{r}_e,\mathbf{r}_h)=-\frac{e^2}{\epsilon |\mathbf{r}_e-\mathbf{r}_h|},
\end{equation}
is the potential energy of the Coulomb interaction between electron and hole, and $\epsilon$ is the permittivity. The Schr\"odinger equation with the periodically time-dependent Hamiltonian (\ref{HEMW}) corresponds to the Floquet problem for an exciton dressed by the field (\ref{A}). To simplify the problem, let us apply the the Kramers-Henneberger unitary transformation~\cite{Kramers,henneberger1968perturbation},
\begin{equation}\label{K}
\hat{\mathcal{\cal U}}(t)=\exp\hspace{-0.2em}{\left[\frac{i}{\hbar c}\sum_{j=e,h}\frac{e}{m_j}\int^t \hspace{-0.5em} dt' \left( \mathbf{A}(t')\xi_j\hat{\mathbf{p}}_j-\frac{eA^2(t')}{2c}\right)\right]},
\end{equation}
which removes the coupling of the momentums $\hat{\mathbf{p}}_{e,h}$ to the vector potential $\mathbf{A}(t)$ in the Hamiltonian (\ref{HEMW}) and transfers the time dependence from the kinetic energy of electron and hole to their potential energy (\ref{UC}) [see Appendix A for details].
Then the transformed Hamiltonian (\ref{HEMW}) reads
\begin{eqnarray}\label{Ht}
\hat{\cal H}&=&\hat{\cal U}^\dagger(t)\hat{\cal H}_X\hat{\cal U}(t) -
i\hbar\hat{\cal U}^\dagger(t)\partial_t
\hat{\cal U}(t)\nonumber\\
&=&\sum_{j=e,h}\frac{\hat{\mathbf{p}}_j^2}{2m_j}
+U_C(\mathbf{r}_e-\mathbf{r}_{e}^\prime(t),\mathbf{r}_h-\mathbf{r}_{h}^\prime(t)),
\end{eqnarray}
where
\begin{equation}\label{rR}
\mathbf{r}_{e(h)}^\prime(t)=\xi_{e(h)}\bar{r}_{e(h)}(\cos\omega_0 t, -\sin\omega_0 t)
\end{equation}
is the radius-vector describing the classical circular trajectory of electron (hole) in the field (\ref{A}), and $\bar{r}_{e(h)}=|e|E_0/m_{e(h)}\omega_0^2$ is the radius of the trajectory.

The Hamiltonian (\ref{Ht}) is still physically equal to the exact Hamiltonian of irradiated exciton (\ref{HEMW}). To proceed, we have to make the approximation described in Appendix B. Namely, considering the Floquet problem with the Hamiltonian (\ref{Ht}), one can apply the $1/\omega_0$ expansion (the Floquet-Magnus expansion~\cite{Eckardt_2015,Goldman_2014,Bukov_2015,Casas_2001}) in order to turn the periodically time-dependent Hamiltonian (\ref{Ht}) into the effective stationary Hamiltonian
\begin{equation}\label{Hef}
\hat{\cal H}_{\mathrm{eff}}=\hat{\cal H}_{0}+\sum_{n=1}^\infty\frac{[\hat{\cal H}_{n},\hat{\cal H}_{-n}]}{n\hbar\omega_0}+{\it O}\left[\left(\frac{\hat{\cal H}_{n}}{\hbar\omega_0}\right)^2\right],
\end{equation}
where $\hat{\cal H}_{n}$ are the components of the Fourier expansion of the time-dependent Hamiltonian (\ref{Ht}), $\hat{\cal H}=\sum_{n=-\infty}^{\infty}\hat{\cal H}_ne^{in\omega_0 t}$. Assuming the field frequency $\omega_0$ to be high enough, one can restrict the Floquet-Magnus expansion (\ref{Hef}) by the term $\hat{\cal H}_{0}$. Then the effective Hamiltonian (\ref{Hef}) reads
\begin{equation}\label{Heff}
\hat{\cal
H}_{\mathrm{eff}}=\sum_{j=e,h}\frac{\hat{\mathbf{p}}_j^2}{2m_j}
+U_0(\mathbf{r}_e,\mathbf{r}_h),
\end{equation}
where
\begin{equation}\label{U0}
U_0(\mathbf{r}_e,\mathbf{r}_h)=\frac{1}{2\pi}\int_{-\pi}^{\pi}U_C(\mathbf{r}_e-\mathbf{r}_{e}^\prime(t),\mathbf{r}_h-\mathbf{r}_{h}^\prime(t))d(\omega
t)
\end{equation}
is the Coulomb potential (\ref{UC}) modified by the dressing field (\ref{A}) (the dressed Coulomb potential). Introducing the new coordinates $\mathbf{R}=(m_e\mathbf{r}_e+m_h\mathbf{r}_h)/(m_e+m_h)$ and ${\mathbf{r}}=\mathbf{r}_e-\mathbf{r}_h$, the effective Hamiltonian (\ref{Heff}) can be rewritten as
\begin{equation}\label{Hefff}
\hat{\cal H}_{\mathrm{eff}}=\frac{\hat{\mathbf{P}}^2}{2M}+\frac{\hat{\mathbf{p}}^2}{2\mu}+U_0({r}),
\end{equation}
where $\mathbf{R}$ is the radius-vector of the exciton center of mass,  ${\mathbf{r}}$ is the radius-vector of relative position of electron and hole, $M=m_e+m_h$ is the total exciton effective mass, $\mu=m_em_h/(m_e+m_h)$ is the reduced exciton mass, $\hat{\mathbf{P}}=-i\hbar\nabla_{\mathbf{R}}$ and $\hat{\mathbf{p}}=-i\hbar\nabla_{\mathbf{r}}$ are the momentum operators in the new coordinates,
\begin{eqnarray}\label{U00}
U_0({r})&=&
\left\{\begin{array}{rl}
-({2e^2}/{\pi r_0\epsilon})K\left({r}/{r_0}\right),
&{r}/{r_0}\leq1\\\\
-({2e^2}/{\pi r\epsilon})K\left({r_0}/{r}\right),
&{r}/{r_0}\geq1
\end{array}\right.
\end{eqnarray}
is the dressed Coulomb potential (\ref{U0}) written in an explicit form (see Fig.~2), the function $\mathrm{K}(\zeta)$ is the elliptical integral of the first kind, and
\begin{equation}\label{r0}
r_0=\frac{|e|E_0}{\mu\omega_0^2}
\end{equation}
is the radius of the classical circular trajectory in the circularly polarized field (\ref{A}) for a particle with the mass $\mu$ and the charge $e$. Thus, the Floquet problem for excitons in the QW irradiated by the dressing field (\ref{A}) is reduced to the Schr\"odinger problem with the stationary Hamiltonian (\ref{Hefff}), which is under consideration below.
\begin{figure}[h!]
\centering\includegraphics[width=0.7\columnwidth]{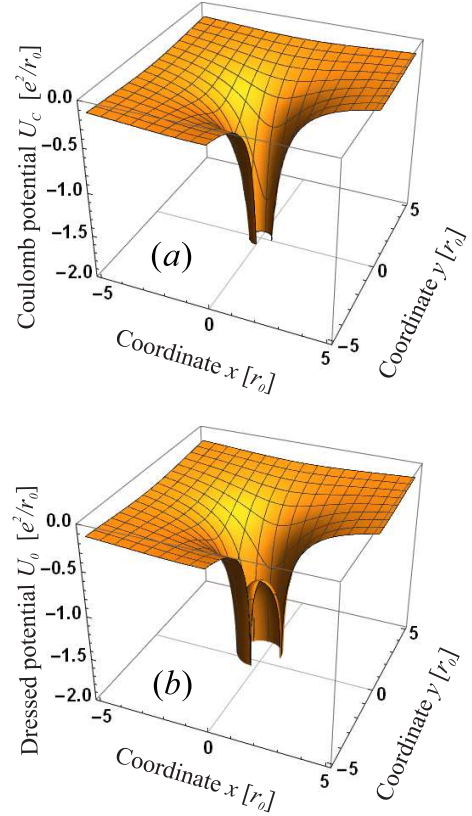}
\caption{(a) The bare Coulomb potential $U_C(r)$  and (b) the dressed Coulomb potential $U_0(r)$  plotted as functions of the planar coordinates $x,y$.} \label{fig2}
\end{figure}

\section{Results and discussion}
The eigenfunction of the Hamiltonian (\ref{Hefff}) reads
\begin{equation}\label{eF}
\Psi(\mathbf{R},\mathbf{r})=\frac{1}{\sqrt{2\pi S}}e^{i\mathbf{K}\mathbf{R}}e^{im\phi}\psi_X(r),
\end{equation}
where $\mathbf{K}$ is the wave vector of the exciton center of mass, $m$ is the exciton angular momentum, $\varphi$ is the polar angle in the $x,y$ plane, $S$ is the area of the QW, and the wave function $\psi_X(r)$ defines the radiative broadening of the exciton spectral line~\cite{Fang_2019}
\begin{equation}\label{G}
\Gamma_X=\frac{2\pi e^2 |p_{cv}|^2}{cm_0^2\omega_X\sqrt{\epsilon}}|\psi_X(0)|^2,
\end{equation}
where $p_{cv}$ is the interband matrix element of the momentum for the semiconductor, $m_0$ is the mass of electron in vacuum, $\hbar\omega_X=\varepsilon_g-\varepsilon_X$ is the photon energy of the exciton emission, $\varepsilon_g$ is the semiconductor band gap, and $\varepsilon_X$ is the exciton binding energy. In the following, we will restrict the analysis by the exciton states with the angular momentum $m=0$, which make the main contribution to the exciton emission. Indeed, it follows from the Hamiltonian (\ref{Hefff}) that $\psi_X(r)\propto r^{|m|}$ near $r=0$ and, therefore, the radiative lifetime of exciton is $\tau_X=\hbar/2\Gamma_X\rightarrow\infty$ for $m\neq0$. Then we arrive from the Hamiltonian (\ref{Hefff}) to the Schr\"odinger equation,
\begin{equation}\label{Sr}
\left[-\frac{\hbar^2}{2\mu}\frac{1}{r}\frac{\partial}{\partial r}\left(r\frac{\partial}{\partial r}\right)+U_0(r)\right]\psi_X(r)
={\cal E}_X\psi_X(r),
\end{equation}
which defines both the exciton wave function $\psi_X(r)$ and the exciton binding energy $\varepsilon_X=-{\cal E}_X$.

The solutions of the Schr\"odinger equation (\ref{Sr}) depend on the two characteristic scales: The exciton Bohr radius
\begin{equation}\label{aX}
a_X=\frac{\hbar^2\epsilon}{\mu e^2}
\end{equation}
and the field-induced trajectory radius $r_0$ described by Eq.~(\ref{r0}). In what follows, we will consider the case of weak dressing field (\ref{A}), when the condition $a_X\gg r_0$ meets. Under this condition, the dressed potential (\ref{U00}) can be considered as a weakly perturbed Coulomb potential (\ref{UC}) and, therefore, the conventional perturbation theory can be applied to solve the Schr\"odinger equation (\ref{Sr}). In the first order of the perturbation theory,
the exciton wave function, $\psi_X(r)$, and the corresponding exciton binding energy, $\varepsilon_X$, are described by the equations
\begin{equation}\label{psiX0}
\psi_X^{(n)}(r)=\psi_{X0}^{(n)}(r)-\sum_{l\neq n}\frac{\langle\psi_{X0}^{(l)}(r)|\Delta U(r)|\psi_{X0}^{(n)}(r)\rangle}
{\varepsilon_{X0}^{(l)}-\varepsilon_{X0}^{(n)}}\psi_{X0}^{(l)}(r)
\end{equation}
and
\begin{equation}\label{eX0}
\varepsilon_X^{(n)}=\varepsilon_{X0}^{(n)}-\langle\psi_{X0}^{(n)}(r)|\Delta U(r)|\langle\psi_{X0}^{(n)}(r)\rangle,
\end{equation}
respectively, where $\Delta U(r)=U_0(r)-U_C(r)$ is the perturbation potential and the index $n$ labels different eigenstates of the Schr\"odinger problem (\ref{Sr}). Here $\psi_{X0}^{(n)}(r)$ and $\varepsilon_{X0}^{(n)}$ are the solutions of the problem in the absence of irradiation $(E_0=0)$, when the dressed potential (\ref{U0}) turns into the usual two-dimensional Coulomb potential (\ref{UC}). In what follows, we will consider only the ground exciton state, omitting the index $n$ in both the wave function (\ref{psiX0}) and the binding energy (\ref{eX0}). We will also calculate the wave function (\ref{psiX0}) only for the zero coordinate, $r=0$, since its value at this point, $\psi_X(0)$, defines the broadening (\ref{G}). Substituting the known solutions of the Schr\"odinger problem for the two-dimensional Coulomb potential~\cite{Yang_1991} into Eqs.~(\ref{psiX0}) and (\ref{eX0}), after straightforward but cumbersome calculations we arrive at the simple equations
\begin{eqnarray}\label{psiX}
\psi_X(0)&=&\frac{4}{a_X}\left[1- 8\left(1-\frac{\pi^2}{16}-\ln{\sqrt{2}}\right)\left(\frac{r_0}{a_X}\right)^2\right]
\nonumber\\
&+&{\it O}\left[\left(\frac{r_0}{a_X}\right)^4\right]
\end{eqnarray}
and
\begin{equation}\label{eX}
\varepsilon_X=\frac{2e^2}{a_X}\left[1-8\left(\frac{r_0}{a_X}\right)^2\right]
+{\it O}\left[\left(\frac{r_0}{a_X}\right)^4\right].
\end{equation}

For definiteness, let us restrict the following analysis by a GaAs-based QW with the band gap $\varepsilon_g=1.51~\mathrm{eV}$, the permittivity $\epsilon=12$, and the effective masses of electrons and holes $m_e=0.067m_0$ and $m_h=0.47m_0$, respectively. The calculation results are presented in Fig.~3 for the different irradiation intensities $I=cE_0^2/4\pi$ and field frequencies $\omega_0$.
Comparing the dressed Coulomb potential (\ref{U0}) and the ``bare'' Coulomb potential (\ref{UC}) plotted in Fig.~2, one can conclude that the dressing field (\ref{A}) induces the repulsive area in the dressed potential $U_0(r)$ near $r=0$ with the local maximum at $r=0$, shifting the attractive potential well from the point $r=0$ to the ring of the radius $r=r_0$. As a result, this repulsive area decreases the exciton binding energy, $\varepsilon_X$. Correspondingly, the field-induced shift of photon energy of the exciton emission is $\hbar\Delta\omega_X=\hbar(\omega_X-\omega_{X0})=-(\varepsilon_X-\varepsilon_{X0})>0$, where $\omega_{X0}$ and $\varepsilon_{X0}$ are the frequency of exciton emission and the exciton binding energy in the absence of irradiation, respectively. Thus, the irradiation results in the blue shift of the exciton emission, which increases with increasing the irradiation intensity $I$ (see Fig.~3a). The same field-induced repulsive area of the dressed potential $U_0(r)$ increases the distance between electron and hole and, therefore, decreases the exciton wave function value (\ref{psiX}). As a consequence, the irradiation decreases the radiative broadening ($\ref{G}$) plotted in Fig.~3b, what means the field-induced narrowing of the exciton spectral line. It should be noted that the decreasing of the broadening $\Gamma_X$ leads also to increasing the exciton lifetime $\tau_X=\hbar/2\Gamma_X$, i.e. the field-induced stabilization of exciton (dynamical stabilization) appears. This looks promising to control the quantum effects originated from the Bose nature of excitons (particularly, the Bose-Einstein condensation of them~\cite{Butov_1994,Butov_2002}) which strongly depend on the lifetime.

\begin{figure}[h!]
\centering\includegraphics[width=0.8\columnwidth]{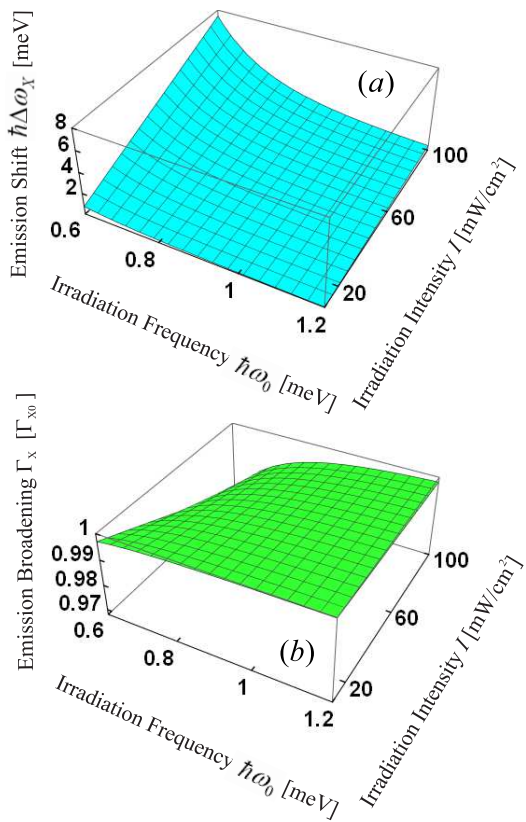}
\caption{(a) The irradiation-induced shift of the light emission frequency, $\Delta\omega_X$, from the ground exciton state in the GaAs-based QW as a function of the irradiation intensity, $I$, and the irradiation frequency, $\omega_0$; (b) The radiative broadening of the exciton spectral line, $\Gamma_X$, as a function of the irradiation intensity, $I$, and the irradiation frequency, $\omega_0$ for the same QW, where $\Gamma_{X0}$ is the broadening in the absence of irradiation.}\label{fig3}
\end{figure}

To discuss the coordinate dependence of the eigenfunctions of the Hamiltonian (\ref{Hefff}), it should be noted that the dressed potential (\ref{U00}) turns into the usual Coulomb potential (\ref{UC}) in the absence of irradiation. In this case, the discussed exciton eigenfunction coincides with the well-known eigenfunction of two-dimensional hydrogen atom (which is the product of the exponent and the confluent hypergeometric function)~\cite{Yang_1991}. Particularly, the eigenfunction of the ground exciton state in the absence of irradiation reads $\psi_{X0}(r)=(4/a_X)e^{-2r/a_X}$. It follows from the aforesaid that the irradiation perturbs the Coulomb potential $U_C(r)$ substantially only near its origin point $r=0$, inducing the local potential maximum there (see Fig.~2). Therefore, the eigenfunctions of the Hamiltonian (\ref{Hefff}) in the presence of the irradiation are the above-mentioned eigenfunctions of two-dimensional hydrogen atom~\cite{Yang_1991}, which are slightly perturbed (reduced) by the irradiation near $r=0$.

To discuss the polarization dependence of the excitonic effects, it should be noted that the dressed Coulomb potential $U_0(\mathbf{r})$ crucially depends on the light polarization~\cite{Kibis_2019}. Namely, the circular polarization leads to the local maximum of the potential at $r=0$ (see Fig.~2b), whereas the linear polarization induces only the saddle point at $r=0$ (see Fig.~2a in Ref.~\onlinecite{Kibis_2019}). Since the discussed excitonic effects originate physically from the light-induced electron-hole repulsion near $r=0$, they are more pronounced for the dressed potential with the local maximum which repulses an electron and a hole from each other more effectively than the saddle-like potential. Therefore, the circular polarization considered in the present article is preferred to observe the discussed excitonic effects experimentally.

The present theory is developed under the assumption of stable exciton states. Therefore, it describes the problem accurately if the exciton lifetime, $\tau_X$, much exceeds the dressing field period, $T$, i.e. the condition $\omega_0\tau_X\gg1$ should be satisfied. In semiconductor QWs fabricated with using the modern nanotechnologies, the lifetime $\tau_X$ is of tens picosecond scale and, therefore, the developed theory is applicable to dressing field frequencies $\omega_0$ starting from the upper microwave range.

Finally, one has to stress again that the discussed excitonic effects appear due to the feature of the dressed Coulomb potential near $r=0$, where the dressing field (\ref{A}) induces the repulsive area (see Fig.~2b). This differs crucially from the case of the repulsive Coulomb interaction between charge particles. Indeed, the circularly polarized field (\ref{A}) applied to the repulsive Coulomb potential induces the {\it attractive} area in the core of the potential, which leads to the electron states bound at the repulsive potential and, particularly, to the field-induced electron pairing~\cite{Kibis_2019,Kibis_2020_2,Kibis_2021_2,Kibis_2021_3,Iorsh_2021}.

\section{Conclusion}
We showed theoretically that a high-frequency off-resonant electromagnetic field can serve as a tool to control excitons in semiconductor quantum wells. Since the excitons form the main mechanism for light emission from semiconductor structures, the developed theory can find its application to tune various characteristics of the optoelectronic devices based on quantum wells, including the emission frequency and the radiative broadening of the spectral line. Moreover, it can be used also to control various effects depending on the lifetime of excitons, which rises under the irradiation. Since the onset of field-induced excitonic effects happens at the timescale of the field period, the discussed method to control excitonic properties of quantum wells by a high-frequency electromagnetic field is very fast as compared to the relatively slow electrostatic control of them by the gate voltage. This is the doubtless advantage from viewpoint of its possible applications to the fast optoelectronic devices based on quantum wells. The present analysis is performed for conventional GaAs-based semiconductor quantum wells and demonstrates that the discussed effects can be observable for relatively weak irradiation with the intensity of tens mW/cm$^2$.

\section{Acknowledgements}
The reported study was funded by the Russian Science Foundation (project 20-12-00001).

\appendix

\section{The Kramers-Henneberger unitary transformation}
The Kramers-Henneberger unitary transformation~\cite{Kramers,henneberger1968perturbation},
\begin{equation}\label{KA}
\hat{\cal U}(t)=\exp\left\{\frac{i}{\hbar}\int^{\,t}\left[
\frac{e}{mc}\mathbf{A}(\tau)\hat{\mathbf{p}}-\frac{e^2}{2mc^2}A^2(\tau)
\right]d\tau\right\},
\end{equation}
was introduced into the quantum mechanics in order to remove the coupling of the momentum $\hat{\mathbf{p}}$ to the vector potential $\mathbf{A}(t)$ in the conventional Hamiltonian,
\begin{equation}\label{HA}
\hat{\cal H}=\frac{[\hat{\mathbf{p}}-e\mathbf{A}(t)/c]^2}{2m}+U(\mathbf{r}),
\end{equation}
describing the behaviour of a particle with the charge $e$ and the mass $m$ in the electromagnetic field with the vector potential $\mathbf{A}(t)$ and the potential energy $U(\mathbf{r})$. Taking into account that
\begin{equation}\label{TA}
\hat{T}=\exp\left[-\frac{ie}{\hbar mc}\int^{\,t}
\mathbf{A}(\tau)\hat{\mathbf{p}}d\tau\right]
\end{equation}
is the translation operator, we arrive at the equality~\cite{henneberger1968perturbation}
\begin{equation}\label{fA}
\hat{T}f(\mathbf{r})=f\left(\mathbf{r}-\frac{e}{mc}\int^{\,t}
\mathbf{A}(\tau)d\tau\right),
\end{equation}
where $f(\mathbf{r})$ is any function depending on the particle coordinate $\mathbf{r}$. It follows from Eq.~(\ref{fA}) that the applying of the unitary transformation (\ref{KA}) to the Hamiltonian (\ref{HA}) results in the Hamiltonian
\begin{eqnarray}\label{HtA}
\hat{\cal H}^\prime&=&\hat{\cal U}^\dagger(t)\hat{\cal H}\hat{\cal U}(t) -
i\hbar\hat{\cal U}^\dagger(t)\partial_t
\hat{\cal U}(t)\nonumber\\
&=&\frac{\hat{\mathbf{p}}^2}{2m}+U\left(\mathbf{r}-\frac{e}{mc}\int^{\,t}
\mathbf{A}(\tau)d\tau\right),
\end{eqnarray}
which is physically equal to the initial Hamiltonian (\ref{HA}). However, in contrast to the Hamiltonian (\ref{HA}), the terms $\propto\hat{\mathbf{p}}$ are absent in the Hamiltonian (\ref{HtA}), what simplifies solving many Schr\"odinger problems. Taking into account the different masses and charges of electrons and holes, we arrive easily from Eq.~(\ref{KA}) to the unitary transformation (\ref{K}) corresponding to the considered exciton problem.

\section{The Floquet-Magnus expansion}
In the most general form, the nonstationary Schr\"odinger equation
for an electron in a periodically time-dependent field with the
frequency $\omega_0$ can be written as
$i\hbar\partial_t\psi(t)=\hat{\cal H}(t)\psi(t)$, where $\hat{\cal
H}(t+T)=\hat{\cal H}(t)$ is the periodically time-dependent
Hamiltonian and $T=2\pi/\omega_0$ is the field period. It follows
from the well-known Floquet theorem that solution of the Schr\"odinger
equation is the Floquet function, $\psi(t)=e^{-i\varepsilon
t/\hbar}\varphi(t)$, where $\varphi(t+T)=\varphi(t)$ is the
periodically time-dependent function and $\varepsilon$ is the
electron (quasi)energy describing behavior of the electron in the
periodical field. The Floquet problem is aimed to find the electron energy spectrum, $\varepsilon$. To solve the problem, let us introduce the unitary transformation, $\hat{\cal U}=e^{iS}$, which transfers the time dependence from the Hamiltonian $\hat{\cal H}(t)$ to its basis states. Then we arrive from the time-dependent Hamiltonian $\hat{\cal H}(t)$  to the effective time-independent Hamiltonian
\begin{equation}\label{HefA}
\hat{\cal H}_{\mathrm{eff}}=\hat{\cal U}^\dagger\hat{\cal H}\hat{\cal U} -
i\hbar\hat{\cal U}^\dagger\partial_t
\hat{\cal U}.
\end{equation}
Solving the stationary Schr\"odinger problem with the Hamiltonian (\ref{HefA}), $\hat{\cal H}_{\mathrm{eff}}\Psi=\varepsilon\Psi$, one can find the sought electron energy spectrum, $\varepsilon$.

There is the regular method to find the transformation matrix $S$ in the particular case of high field frequency, $\omega_0$, which satisfies the condition ${\hat{\cal H}_{n}}/{\hbar\omega_0}\ll1$, where $\hat{\cal H}_{n}$ are the components of the Fourier expansion of the time-dependent Hamiltonian, $\hat{\cal H}(t)=\sum_{n=-\infty}^{\infty}\hat{\cal H}_ne^{in\omega_0 t}$. Namely, both the transformation matrix $S$ and the effective Hamiltonians (\ref{HefA}) can be found as the $1/\omega_0$ expansion (the Floquet-Magnus expansion)~\cite{Eckardt_2015,Goldman_2014,Bukov_2015,Casas_2001}. As a result, we arrive at the effective stationary Hamiltonian
(\ref{Heff}). Thus, the Floquet-Magnus expansion reduces the time-dependent Floquet problem to the conventional stationary Schr\"odinger problem with the time-independent Hamiltonian (\ref{Heff}).

\end{document}